\documentstyle[10pt]{article}

\def\bfl{\begin{flushleft}}
\def\efl{\end{flushleft}}
\def\bfr{\begin{flushright}}
\def\efr{\end{flushright}}
\def\bc{\begin{center}}
\def\ec{\end{center}}
\def\be{\begin{equation}}
\def\ee{\end{equation}}
\def\ba{\begin{eqnarray}}
\def\ea{\end{eqnarray}}
\def\nn{\nonumber }

\def\text#1{\mbox{#1}}
\def\drm{\text{d}}
\def\tableline{\hline}
\def\Sign#1{\, \text{sign}\left[#1\right] }
\def\Eq{Eq.~}
\def\Eqs{Eqs.~}
\def\Ref{Ref.~}
\def\Het{$^3\text{He}$ }
\def\Hef{$^4\text{He}$}
\def\schrod{Schr\"oedinger  }

\begin{document}

~\\
\bfr
Acta Phys. Polon. B 30 (No.4) pp. 897-905 (1999)\\
gr-qc/9802060\\
\efr
~\\
~\\
~\\
~\\
\bc
{\LARGE \bf
Acoustic phase lenses in superfluid helium as models of composite spacetimes
in general relativity: Classical and quantum features
}

~~\\
{\large Konstantin G. Zloshchastiev}\\
~~\\
Box 2837, Dnepropetrovsk 320128, Ukraine;                             
e-mail:  zlosh@email.com  \\
~\\
Received: 24 Feb 1998 (LANL), 26 Oct 1998 (APPB)  
\ec

~~\\

\abstract{\large
In the spirit of the well-known analogy between inviscid fluids and
pseudo-Riemannian manifolds we study spherical singular hypersurfaces
in the static superfluid.
Such hypersurfaces turn out to be the interfaces dividing the superfluid 
into the
pairs of spherical domains, examples of which are 
``superfluid A - superfluid B'' or 
``impurity - superfluid'' phases.
It is shown that these shells form the acoustic lenses which are the
sonic counterparts of the usual optical ones. 
The exact equations of motion of the lens interfaces are obtained.
Also some quantum aspects of the theory
are considered.
We calculate energy 
spectra for bound states of acoustic lenses in dynamical equilibrium, 
taking into account the analogy to a material shell model of a 
black hole (we consider the cases of spatial topology of a black hole 
and a wormhole type).
}

~\\

PACS number(s):  04.40.Nr, 11.27.+d, 43.35.+d, 67.57.De\\

~~\\
~~\\
~~\\
~~\\
~~\\

\newpage
\large

It was shown in numerous works that the superfluid phases of \Het 
(and perhaps \Hef) can simulate
phenomena encountered in gravitation and the standard model of 
elementary particles.
Physics of superfluid \Het illustrates concepts in
quantum field theory and gravity such as:
black holes, surface gravity, Hawking radiation, horizons, ergoregions,
trapped surfaces
\cite{unr1,jac,unr2} (see \Ref \cite{vis} for an introduction into recent 
developments), 
baryogenesis, vortexes, strings, textures, standard electroweak model 
(see \cite{vv,vol,vol-jetp}, and references therein), and so on.
This turns out to be possible due to the certain analogy between 
inviscid fluids and pseudo-Riemannian manifolds.
The simplest way to show this correspondence is as follows.

The fundamental equations of dynamics of an inviscid fluid are the 
Euler equation
\be
\rho
\left[
      \frac{\partial \vec v}{\partial t} +
      (\vec v \cdot \nabla) \vec v
\right] = - \nabla p - \rho \nabla \Phi,                  \label{eq1}
\ee
and equation of continuity
\be
\frac{\partial \vec v}{\partial t} +
\nabla\cdot(\rho \vec v) = 0,                             \label{eq2}
\ee
where $\Phi$ is the potential of an external force (including gravity),
$\vec v$ is the flow velocity, $\rho$ and $p$ are, respectively,
the fluid density and pressure.
If one assumes the flow to be locally irrotational then we can introduce
the velocity potential $\psi$, $\vec v = -\nabla\psi$.
Hence, assuming the barotropic equation of state $\rho (p)$, 
the Euler equation 
can be rewritten in the form of the Bernoulli equation \cite{vis}
\be
-\frac{\partial \psi}{\partial t} 
+\frac{1}{2} (\nabla \psi)^2
+\int\limits_{0}^{p} \frac{\drm p^\prime}{\rho (p^\prime)}
+\Phi = 0.                                                 \label{eq3}
\ee
We can linearize these equations around some background 
$\{ \rho_0, p_0, \psi_0 \}$ to consider the propagation of small 
fluctuations (sound waves).
We suppose 
$\rho=\rho_0+\epsilon \rho_1 + o (\epsilon)$, 
$p=p_0+\epsilon p_1 + o (\epsilon)$, 
$\psi=\psi_0+\epsilon \psi_1 + o (\epsilon)$,
the external potential is fixed.
Then, linearizing the Euler equation and taking into account the linearized
continuity equation, we finally obtain the wave equation describing
the propagation of the fluctuation $\psi_1$
\ba
&&-\frac{\partial }{\partial t}
\left[
\rho_0
\frac{\partial \rho}{\partial p}
\left(
\frac{\partial \psi_1}{\partial t}
+\vec v_0 \cdot \nabla \psi_1
\right) 
\right]             \nn\\
&&+ \nabla \cdot
\left[
\rho_0 \nabla \psi_1
-\rho_0 \vec v_0 \frac{\partial \rho}{\partial p}
\left(
\frac{\partial \psi_1}{\partial t}
+\vec v_0 \cdot \nabla \psi_1
\right) 
\right]  = 0.                                                \label{eq4}
\ea
This equation can be rewritten as the d'Alembert equation in the
curved background spacetime
\be
\frac{1}{\sqrt{-g}}
\frac{\partial }{\partial x^\mu}
\left(
\sqrt{-g}~ g^{\mu\nu}
\frac{\partial \psi_1}{\partial x^\nu}
\right) = 0,                                                \label{eq5}
\ee
where $\mu = \{0, i\}$, $x^\mu = \{t, \vec x\}$, $g=\det{(g_{\mu\nu})}$, 
and the acoustic background metric is
\be
\drm s^2 = \frac{\rho_0}{c} 
\left[
-c^2 \drm t^2 + 
\delta_{i j} (\drm x^i - v_0^i \drm t)  (\drm x^j - v_0^j \drm t)
\right],                                                    \label{eq6}
\ee
where $c = \sqrt{\partial p / \partial \rho}$ is the local speed of sound.
Thus, the vorticity-free flow of a zero viscosity fluid can 
be seen to define a Lorentzian signature metric a curved space-time.
Of course, the physical space-time is just the usual Minkowski flat 
space-time.

The aim of this paper is to study infinitely thin shells in such
acoustic space-times
as models of physical objects whose thickness is negligible in 
comparison with a  circumference radius
(e.g., surfaces of phase domains).
A thin shell is thought to be a discontinuity of the second kind
(the density has the delta-like singularity on the shell).
Its dynamics
is determined by the Lichnerowicz-Darmois-Israel junction conditions:
the first quadratic form (metric) is continuous, 
the second quadratic form (extrinsic curvature) has a finite 
jump across the shell.
Geometrically a shell is described by a three-dimensional closed 
singular hypersurface, embedded in
the four-dimensional space-time and dividing it into two domains: 
the external ($\Sigma^+$) and internal ($\Sigma^-$) regions of
space-times.
Since the classic works \cite{dau,isr} the theory 
of singular hypersurfaces has been widely considered in the literature 
(see \Ref \cite{mtw} for details).
We describe only some basic properties of timelike hypersurfaces
corresponding to dynamical evolution of thin shells now.
One considers a singular matter layer 
$\Sigma$ described by the three-dimensional 
space-time with the surface stress-energy 
tensor of a perfect fluid in the general case 
\be
S_{ab}=\sigma u_a u_b 
+ \frac{p}{c_\Sigma^2} (u_a u_b +~ ^{(3)}\!g_{ab}),               \label{eq7}
\ee
where $\sigma$ and $p$ are the surface mass-energy density and 
pressure respectively, $u^a$ is the timelike unit tangent vector, 
$^{(3)}\!g_{ab}$ is the 3-metric of a shell's surface (in the acoustic 
sense (\ref{eq6})),
$c_\Sigma$ is the speed of sound in the shell.
We suppose metrics of the fluid space-times outside $\Sigma^+$ and inside 
$\Sigma^-$ of a spherical shell to be flat:
\be
\drm s_\pm^2 =
-c_{\pm}^2 \drm t^2 +  \drm r^2 + r^2 \drm \Omega^2,            \label{eq8}
\ee
where $\drm \Omega^2$ is the metric of the unit 2-sphere,
$c_\pm$ are the constants of the speed of sound in the space-times 
$\Sigma^\pm$.
These metrics correspond to the spherical shell dividing different phase
domains inside the motionless homogeneous superfluid (\ref{eq6}).
It is possible to show that if one uses the shell's proper time  
$\tau$ then the 3-metric of the shell's space-time history is
\be
^{(3)}\!\drm s^2 = 
-c_\Sigma^2 \drm \tau^2 + R^2 \drm \Omega^2,                \label{eq9}
\ee
where $R(\tau)$ is the shell's radius.
As it can be seen from \Eqs (\ref{eq8}), (\ref{eq9}), 
we obtain the composite space-time consisting of three regions, 
$\Sigma_+$, $\Sigma_-$, and
$\Sigma$, characterizing by proper fundamental constants.
The space-time domains inside and outside a shell (\ref{eq8}) 
are flat\footnote{\normalsize The
presence of two flat spacetimes which nevertheless are regarded
as different does not contradict to the relativity principle 
(which means that metrics are equivalent if one can be transformed
to another by virtue of coordinate transformations) because
here we have three independent domains such that the relativity principle
is satisfied inside of every of them {\it separately}.
Besides, it should be pointed out that such a three-regional union
spacetime does not appear to be a manifold in the wide-used sense
of this word.}
and are characterized only by
the fundamental constants of the speed of sound, 
whereas the three-dimensional shell's space-time can be curved 
(we mean the curvature in the acoustic sense) and, in addition 
to the constant $c_\Sigma$, the ``gravitational'' constant
$\gamma_\Sigma$ may appear as well.
The energy conservation law for a shell 
(which is the shell's interpretation of the integrability 
condition $S^a_{b;a}=0$) can be written as
\be
c_\Sigma^2\, \drm \left( \sigma ~^{(3)}\!g \right) +
p~ \drm \left( ~^{(3)}\!g \right)  =0,                         \label{eq10}
\ee
where  
$^{(3)}\!g=\sqrt{-\det{(^{(3)}\!g_{ab})}} = c_\Sigma R^2 \sin{\theta}$.
In this equation, the first term corresponds to a change in
the shell's internal energy, the second term corresponds to the work done
by the shell's internal forces.

It is important to note that the analogy between inviscid fluids and
pseudo-Riemannian manifolds appears to be justified as yet 
on the kinematical level only.
Thus, the Einstein equations as such have no evident physical sense 
within the frameworks of inviscid fluid dynamics.
Only the (acoustic) metric, the manifold topology, 
and equations of motion (which are the 
consequence of the Bianchi identity) have direct physical interpretation.
However, the above-mentioned junction conditions, strictly speaking,
are connected rigidly neither with general relativity nor with the
Einstein equations, despite the fact that historically they were 
first derived in the contex of general relativity.
They simply represent the procedure of geometrical
matching of two Riemannian manifolds
across a surface of discontinuity of the second kind,
and thus can be supposed independently as equations describing
behaviour of an interface between two liquid phase regions.
In this connection the words of famous mathematician Kolmogorov that
``whole mathematics (therefore, physics too) can be 
reformulated as geometry'' are quite relevant.
By imposing the junction conditions 
\[
(K^a_b)^+ - (K^a_b)^- = 4 \pi\sigma (2 u^a u_b + ~^{(3)}\!\delta^a_b),  
\]
where $(K_{ab})^\pm$ are the extrinsic curvatures of 
the spherically symmetric singular hypersurface \cite{zlo002}
with respect to the external and internal acoustic manifolds $\Sigma^\pm$,
we obtain the equation of shell's motion in the form                                    
\be
\epsilon_+ \sqrt{1+(\dot R/c_+)^2 } - 
\epsilon_- \sqrt{1+(\dot R/c_-)^2 } = - 4 \pi \zeta \sigma R,  \label{eq11}
\ee
where $\dot R=\drm R/\drm\tau$ is the velocity of a shell, 
$\epsilon_\pm = \Sign{\sqrt{1+ (\dot R/c_\pm)^2}}$ (see below), 
$\zeta$ is a fundamental constant for the shell's space-time $\Sigma$
\cite{vol-jetp}, 
$\zeta = \gamma_\Sigma/c_\Sigma^2$ with the dimensionality 
$[\zeta]=\,\text{cm}\,\text{g}^{-1}$.

From \Eq (\ref{eq11}) one can see that we obtain a simple but
nontrivial object.
From the viewpoint of general relativity it has neither mass nor
charge nor some other habitual global property.
The only its global attribute is to be an interface of two space-times with
different fundamental constants.
Nevertheless, such shells have the nontrivial local dynamic 
properties, viz., the proper velocity, tension and mass-energy density 
(therefore, an equation of state).
Moreover, it can easily be seen that the shell's matter 
can be assumed to be enough arbitrary one, both highly exotic and ordinary.   
Once the function $\sigma (R)$ is known then by means of the 
conservation
law (\ref{eq10}) we can obtain the equation of state $p=p(\sigma)$.

It should also be noted that a sound, passing from the one
space-time (\ref{eq8}) to other across the shell-interface, will 
be refracted,
as it happens for light rays in a usual (spherical) lens.
Other analogous phenomena, e.g., the spectral factorization
or focusing of sound, can appear as well.

The equation of motion (\ref{eq11}) together 
with the equation of state (or, equivalently, with the known function
$\sigma (R)$) and choice of the signs $\epsilon_\pm$, completely determines 
the motion of superfluid shells (interfaces of the acoustic lenses).
Therefore, first of all we must specify $\epsilon_\pm$ and $\sigma(R)$.

Let us say few words about the topological features of the theory.
In general relativity it is well-known \cite{bkt,gk} that 
$\epsilon = +1$ if $R$ increases in 
the outward normal direction to the shell,
and $\epsilon = -1$ if $R$ decreases.
Thus, under the  condition $\epsilon_+ = \epsilon_-$ we have the
ordinary (black hole type) shell, and under 
$\epsilon_+ = -\epsilon_-$ we have the traversable 
wormhole type shell \cite{vis2}.
The appropriate cases are represented in the table \ref{tab1} 
(we assume the surface density $\sigma$ to be positive), where 
the shells (i.e., surfaces of the second kind) of ordinary lenses are 
sonic analogs of the black hole 
type shells, and shells corresponding to 
anomalous lenses are counterparts of the 
wormhole type shells \cite{zlo001}.
The superscript ``$\dagger$'' denotes the case of ordinary lenses when
the notions ``outside the shell'' and ``inside the shell'' are reversed 
(for anomalous, wormhole, 
lenses such notions are absent {\it ab initio}).

Below we assume the rate of change of the lens size to be small,
$\dot R \ll c_\pm$.
Otherwise, the disturbances which may occur could be incompatible 
with the assumed flatness of the superfluid space-times (\ref{eq8}).
Following (\ref{eq11}), we obtain the equation of motion of the lenses
in the form
\be
\frac{\mu \dot R^2}{2} = \Xi (R),                            \label{eq12}
\ee
where
\ba
&&\Xi (R) = 4\pi\zeta\sigma R -2 \delta,                   \nn\\
&&\mu=c_-^{-2} + (2\delta - 1) c_+^{-2} > 0,                  \label{eq13}\\
&& \delta=\left\{ \,{}_1^0  \right. \quad 
{}_{\text{(AL)}}^{\text{(OL)}},          \nn
\ea
and we call $\delta$ the {\it parameter of lens anomaly}.

Further, we do not know what is the concrete matter in the shell.
However, we can specify the class of the lenses 
in the dynamical equilibrium
(the other lenses will be either growing or decreasing in size and this 
eventually leads to the vanishing 
of either of the two phases $\Sigma^\pm$).
The Taylor expansion of the function $\Xi (R)$ in a small neighborhood
of the equilibrium point $R_0$ yields
\be
\Xi (R) = -2 \delta + \varepsilon - \frac{k^2}{2} (R-R_0)^2 
+ o\, ((R-R_0)^2),                                              \label{eq14}
\ee
where $\varepsilon$ and $k$ are the constants,
\[
\varepsilon = 4 \pi \zeta \sigma|_{R=R_0} R_0,\>
k^2 =- 4 \pi \zeta (\sigma R)^{\prime\prime}|_{R=R_0}.
\]
Then \Eq (\ref{eq12}) can be written as the energy conservation law 
for the harmonic oscillator.
Performing the shift $x=R-R_0$, we obtain 
\be
E = \frac{P^2}{2 m}
+ \frac{2\delta}{\zeta\mu k} + \frac{m \omega^2 x^2}{2},       \label{eq15}
\ee
where 
\[
P = m \dot R = m \dot x,~~ 
E=\frac{\varepsilon}{\zeta\mu k},~~
m=\frac{1}{\zeta k},~~
\omega = \frac{k}{\sqrt{\mu}}. 
\]
It should be noted that $R\in [0,\,+\infty)$ hence $x\in [-R_0,\,+\infty)$.
This circumstance is very important for further studies,
first of all for analysis of quantum aspects of the theory. 

Below we study the quantum mechanical properties of our lenses.
One can perform the standard procedure of quantization.
Then the conservation law (\ref{eq15}) gives us the stationary \schrod
equation for the spatial wave function $\Psi (x)$
\be
-\frac{\hbar^2}{2 m} \frac{\drm^2 \Psi}{\drm x^2}
+
\left[
-E + \frac{2\delta}{\zeta\mu k} + \frac{m \omega^2}{2} x^2
\right]\Psi=0,                                               \label{eq16}
\ee
or, in the dimensionless form,
\be
\frac{\drm^2 \Psi}{\drm y^2}+
\left(
\varrho - y^2
\right)\Psi=0,                                               \label{eq17}
\ee
where
\[
y=\sqrt{\frac{m\omega}{\hbar}} x,~~
\varrho = \frac{2}{\hbar \omega}
\left(
      E - \frac{2\delta}{\zeta\mu k}
\right).
\] 

However \Eq (\ref{eq17}) is not purely the equation for the quantum 
harmonic oscillator,
because the oscillator's wave functions are defined on the line 
$(-\infty,+\infty)$ whereas in the present case we have both the whole
line and the half-line
$y\in [-R_0\sqrt{m\omega/\hbar},\,+\infty)$.
The analytic continuation of $y$ on the whole axis $(-\infty,+\infty)$
can be correctly explained only for the AL type shells because they are
acoustic wormholes as was mentioned above.
Such a continuation of the spatial coordinate appears to be a
somewhere artificial but necessary technique.
Indeed, in the wormhole case we have matched the two
non-embedded space-times, which both have their own
infinitely distant points.
Then after the continuation one can explicitly discriminate these
spatial infinities from each other by virtue of a sign.
Thus, besides the parameter $\delta$, the ordinary and anomalous lenses
have different topological properties.
Below we distinguish these cases.

(i) {\it Anomalous lenses}.
In this case physics admits the analytic continuation  
$y \in (-\infty,+\infty)$.
For bound states the quantum boundary conditions, 
corresponding to the singular Stourm-Liouville problem,
require $\Psi (+\infty) = \Psi (-\infty) = 0$,
and the normalized solution of \Eq (\ref{eq17}) can be 
expressed by means of the Hermite polynomials $H_n (y)$ \cite{jel}
\be
\Psi (y) = \left( 2^n \sqrt{\pi} n\verb|!|   \right)^{-1/2}
\exp{(-y^2/2)}\, H_n (y),                                   \label{eq18}
\ee
where $n=0,\,1,\,2,...\,$.
The discrete values of energy are $\varrho = 2 n +1$ hence, 
taking into account \Eq (\ref{eq13}),  
\be
E_n = \frac{2}{\zeta k (c_-^{-2} + c_+^{-2})}
+ \frac{\hbar k}{2 \sqrt{c_-^{-2} + c_+^{-2}}} (2 n + 1),     \label{eq19}
\ee
that indeed appears to be the energy of the quantum harmonic oscillator plus
wormhole shift.

(ii) {\it Ordinary lenses}.
In this case it is necessary to solve the \schrod equation (\ref{eq16}) 
on the half-axis 
\[
R\in [0, +\infty)\> \Rightarrow \> 
y\in [-R_0\sqrt{m\omega/\hbar},\,+\infty).
\]
Performing the  transformation $z=y^2$ 
(which works like the baker's transformation \cite{moo}), we obtain that 
$z\in [0, +\infty)$. 
Then \Eq (\ref{eq17}) can be 
written as the confluent hypergeometric equation 
\be
z\, \frac{\drm^2 \varphi}{\drm z^2} + 
\left( \frac{3}{2} - z \right) \frac{\drm \varphi}{\drm z} +
\frac{\varrho-3}{4} \varphi  = 0,                  \label{eq20}
\ee
where $\Psi (z) = \exp{(-z/2)}\, \varphi (z)$.
For bound states the quantum boundary conditions
require $\Psi (0) = \Psi (+\infty) = 0$,
thereby the confluent hypergeometric functions $\varphi (z)$
turn to be the Laguerre polynomials $L_n^{(\alpha)} (z)$, 
$\alpha > -1$ \cite{zlo001,jel}.
Finally we obtain the normalized wave functions
\be
\Psi (y) = \sqrt{\frac{n\verb|!|}{\Gamma (n+1/2)}} y
\exp{(-y^2/2)}\, L_n^{(1/2)} (y^2),                           \label{eq21}
\ee
where $n=0,\,1,\,2,...\,$.
Then we have the observable spectrum of energy to be determined by the
expression
$\varrho = 4 n +3$ 
hence, taking into account \Eq (\ref{eq13}),
\be                                                             \label{eq22}
E_n = \frac{\hbar k}{2 \sqrt{c_-^{-2} - c_+^{-2}}} (4 n +3),    
\ee
which does not depend explicitly on the shell's fundamental 
constant $\zeta$ as it can easily be seen, but involves the 
constant $k$ related to specific matter on the shell.
Comparing expressions (\ref{eq19}) and (\ref{eq22}) we conclude that
\be                                                           \label{eq23}
E^{\text{(AL)}}_{2n+1} - \frac{2}{\zeta k (c_-^{-2} + c_+^{-2})}  = 
E^{\text{(OL)}}_{n},    
\ee
that can be proved also by means of the relation between the 
Laguerre and Hermite polinomials.

In present paper the classical and the quantum aspects of the 
spherically symmetric thin shells in the motionless homogeneous superfluid 
helium were studied.
We have considered such singular hypersurfaces as the traversable 
interfaces between 
pairs of the domains, for instance, the
phases ``$^{3}$He A - $^{3}$He B'', 
the mixtures ``$^{4}$He - $^{3}$He'', or the ``inviscid impurity - He''. 
It was shown that these shells can give rise to the acoustic lenses 
which have to be sonic models of the composite space-time
(i.e., the ``patchwork manifold'' or union space) 
consisting of regions with different fundamental constants.

\def\CMPh{Commun. Math. Phys.}
\def\JPh{J. Phys.}
\def\CJP{Czech. J. Phys.}
\def\FP{Fortschr. Phys.}
\def\LMPh {Lett. Math. Phys.}
\def\MPL  {Mod. Phys. Lett.}
\def\NPh  {Nucl. Phys.}
\def\PhE  {Phys.Essays}
\def\PhL  {Phys. Lett.}
\def\PhR  {Phys. Rev.}
\def\PhRL {Phys. Rev. Lett.}
\def\PhRp {Phys. Rep.}
\def\NCim {Nuovo Cimento}
\def\NuPB {Nucl. Phys.}
\def\GRG {Gen. Relativ. Gravit.}
\def\CQG {Class. Quantum Grav.}
\def\prp {report}
\def\Prp {Report No.}

\def\jn#1#2#3#4#5{{#1}{#2} {\bf #3}, {#4} {(#5)}}  

\def\boo#1#2#3#4#5{{\it #1} ({#2}, {#3}, {#4}){#5}} 

\def\prpr#1#2#3#4#5{{``#1,''} {#2}{#3}{#4}, {#5}} 

\def\And{, and }

\begin{table}                                   
\caption{
The classification of acoustic lenses into the ordinary (OL) and anomalous
(AL) ones, the sign ``$\star$'' denotes 
the impossibility of the Lichnerowicz-Darmois-Israel's junction.}
\bc                 
\begin{tabular}{ccccc}
 \tableline
$\sigma>0$ &\multicolumn{2}{c}{$\epsilon_+ =\epsilon_-$}
 &\multicolumn{2}{c}{$\epsilon_+ =-\epsilon_-$}\\
 &${\epsilon_+=1\choose{\epsilon_-=1}}$ &${\epsilon_+=-
1\choose{\epsilon_-=-1}}$ &${\epsilon_+=1\choose{\epsilon_-
=-1}}$ &${\epsilon_+=-1\choose{\epsilon_-=1}}$\\
 \tableline
 $c_+ > c_-$ &OL&$\star$&$\star$ &AL \\
 $c_+ = c_-$ &$\star$&$\star$&$\star$ &AL \\
 $c_+ < c_-$ &$\star$&OL$^\dagger$&$\star$&AL \\
 \tableline
 \end{tabular}
\ec
 \label{tab1}
 \end{table}

\end{document}